\documentclass[intlimits,twoside,a4paper]{article}

\usepackage[cp1251]{inputenc}
\usepackage{mathrsfs}
\usepackage{multirow}
\usepackage[eqsecnum]{cmpj3}

\issue{2026}{29}{3}{33706}
\doinumber{10.5488/CMP.29.33706}
\title[Light absorption and scattering of an assembly of spheroidal metal nanoparticles]%
{Light absorption and scattering of an assembly of spheroidal metal nanoparticles}
\author[N.~I.~Pavlyshche, A.~V.~Korotun, V.~P.~Kurbatsky]
       {N.~I.~Pavlyshche\orcid{0009-0004-5831-5491}\refaddr{label1},
        A.~V.~Korotun\orcid{0000-0003-4165-2788}\refaddr{label1,label2}\thanks{Corresponding author: \email{andko@zp.edu.ua}.},
        V.~P.~Kurbatsky\orcid{0000-0002-3927-9657}\refaddr{label1}}
\addresses{
\addr{label1} National University Zaporizhzhia Polytechnic, Zaporizhzhia, Ukraine
\addr{label2} G.~V. Kurdyumov Institute for Metal Physics of the NAS of Ukraine, Kyiv, Ukraine
}

\date{Received 12 October 2025; revised 18 August 2026; accepted 25 August 2026; published 28 September 2026}

\begin{document}

\maketitle

\begin{abstract}
The paper studies the optical properties of assemblies of metallic prolate and oblate nanospheroids. The frequency dependences of averaged absorption and scattering cross-sections, absorption, scattering and reflection coefficients are obtained. The calculations were performed for assemblies of spheroidal particles of various metals. The difference in the results of calculating the frequency characteristics of the absorption and scattering cross-sections for assemblies of prolate and oblate spheroids manifested itself in different location and value of the maxima. A qualitative similarity of the frequency dependences of the absorption and scattering coefficients was established. It is shown that the frequency curves of assemblies of spheroids of different metals differ significantly in the location of the extrema within the range from the visible to the ultraviolet region of the spectrum
%
%
\printkeywords
%
\end{abstract}

\section{Introduction}

The composite materials, which contain metallic nanoparticles, are the promising optical materials for use in photonics. For example, the glasses, which contain nanoparticles of plasmonic metals, such as Ag, Au, and Cu, exhibit nonlinear optical properties. They can be used as active media in the ultrafast optical switches \cite{B1}. By altering the shape of the nanoparticles, it is possible to select the spectral region of the optical nonlinearity and control the values of the nonlinear susceptibility. Furthermore, the indicated materials act as metamaterials with the negative permittivity in certain frequency ranges and can be used to create the superlenses \cite{B2,B3,B4,B5}. In the composites, which contain metallic ellipsoids, the nonlinear optical properties are more pronounced than in the composites, which contain spherical nanoparticles \cite{B6,B7,B8,B9}. In optical composite materials, which contain oriented metallic ellipsoids, the nonlinear phenomena such as optical dichroism and birefringence can be observed. These media can also be used as polarizers.

There are a number of methods for synthesizing ellipsoidal silver and gold nanoparticles in liquids, polymers, and on the surfaces \cite{B10}. For example, the paper \cite{B10} describes the obtaining of the prolate ellipsoidal Ag nanoparticles in a polymer. The essence of the method used in this work lies in stretching the silver particles along the direction of the polyimide molecular chain formation during its solidification. Ellipsoidal metallic nanoparticles can also be synthesized in glass and transparent crystals~\cite{B12,B13}. In most cases, however, these ellipsoidal nanoparticles are randomly oriented. The paper~\cite{B14} describes the process for producing oriented silver nanospheroids by stretching the glass containing spherical silver nanoparticles while heating it. These glass samples were stretched to between 50 and 500 times their original length. In \cite{B9,B14+}, oriented silver nanospheres were produced by bombarding the glass containing spherical silver nanoparticles with high-energy ions.

It is known that in the noble metal nanoparticles, one can observe the localized surface plasmon resonance (SPR) of the collective oscillations of conduction electrons induced by external electromagnetic radiation. For this reason, the metallic nanoparticles are used in such fields as surface-enhanced Raman scattering (SERS) \cite{B15,B16}, chemical and biological optical sensors \cite{B17}, solar cells \cite{B18,B19} and light-emitting diodes \cite{B20,B21}. Many of these applications require that the nanoparticles should be uniformly distributed on the substrate or within the dielectric. For example, in modern thin-film solar cells, the requirement for uniform distribution of the plasmonic nanoparticles embedded in the absorber layer is linked to the optimization of the technical characteristics (maximization of efficiency) of such solar cells~\cite{B21+}.

In recent years, a considerable attention has been devoted to the theoretical and experimental study of the light scattering by the metallic micro- and nanoparticles of  different shapes. For example, in the paper \cite{B24}, an approach to studying the dynamic scattering of light by single macroscopic particles was proposed, allowing for a more detailed analysis of the angular scattering characteristics. The review~\cite{B25} systematizes the current understanding of light scattering by small metallic particles and localized surface plasmon resonances, while in the recent study \cite{B26} the influence of the geometric parameters of the metallic nanostructures on the light absorption and scattering characteristics was investigated in detail. These results confirm a decisive role of the shape, size, and dielectric environment of the nanoparticles in determining their optical response.

The work of P. M. Tomchuk, N. I. Grigorchuk, and their co-authors \cite{B27,B28,B29,B30,B31} investigated the influence of the nanoparticle shape and size on the surface plasmon resonance, optical conductivity, light scattering, and  sensitivity of the plasmon modes, as well as took into account the surface effects in the particles of finite size. These works contribute a lot in the development of the theory of the optical properties of the metallic spheroidal nanoparticles. The results obtained in these studies provide an important theoretical foundation for analyzing the optical properties of the ensembles of metallic spheroidal nanoparticles.

The localized SPR peaks of the noble metal nanoparticles are typically found in the visible region of the spectrum or in the near-infrared range. Therefore, many studies investigated the effects of the particle size, film thickness \cite{B32,B33}, coating \cite{B34,B35} and the refractive index of the surrounding medium \cite{B15} on the absorption efficiency and the shift of the peak of localized SPR in metallic nanoparticles. The correlation between the shape of the nanoparticles and the frequency shift of the localized SPR was also reported. These studies primarily used the models with spheroidal nanoparticles \cite{B36,B37}. It was shown that these results are not applicable to the non-spheroidal nanoparticles. A number of studies \cite{B38} considered the particles other than spheroids. In these studies, the statistical averaging by the particle size and shape was performed.

In practical applications, we often do not deal  with the isolated particles, but with the ensembles or composites of the particles. The particles in the ensembles typically vary in the size, shape, and composition, and may randomly change their orientation and position in space. In most cases, it is necessary to know the effective (ensemble-averaged) scattering (absorption) characteristics, such as effective integral cross-sections. Calculating these quantities is a complex task and often requires the use of numerical methods.

The paper \cite{B39} investigated the frequency dependencies of absorption and scattering efficiencies for the ensembles of spherical, cylindrical and disc-shaped metallic nanoparticles, while \cite{B40} studied the frequency dependencies for absorption, scattering, and reflection coefficients of the ensembles of nanodiscs. Although the ensembles of the nanoparticles of deifferent shapes (including spheroidal ones) are used for the optical data storage \cite{B41} and visualization of the cellular structures~\cite{B42}, as well as in biosensing and genomics \cite{B43}, no studies have been conducted on the optical properties of the ensembles of spheroidal metallic nanoparticles. At the same time, most of the published studies focus on individual nanoparticles or regular structures. Meanwhile, for many practical applications, it is necessary to characterize the ensembles of spheroidal metallic nanoparticles in terms of their size and orientation distributions and to determine their effective optical properties. In particular, the calculation of averaged scattering and absorption coefficients for such ensembles remains understudied, which underscores the relevance of this work.
\newpage
\section{Basic relationships}

\begin{figure}[htb]
\centerline{\includegraphics[width=0.85\textwidth]{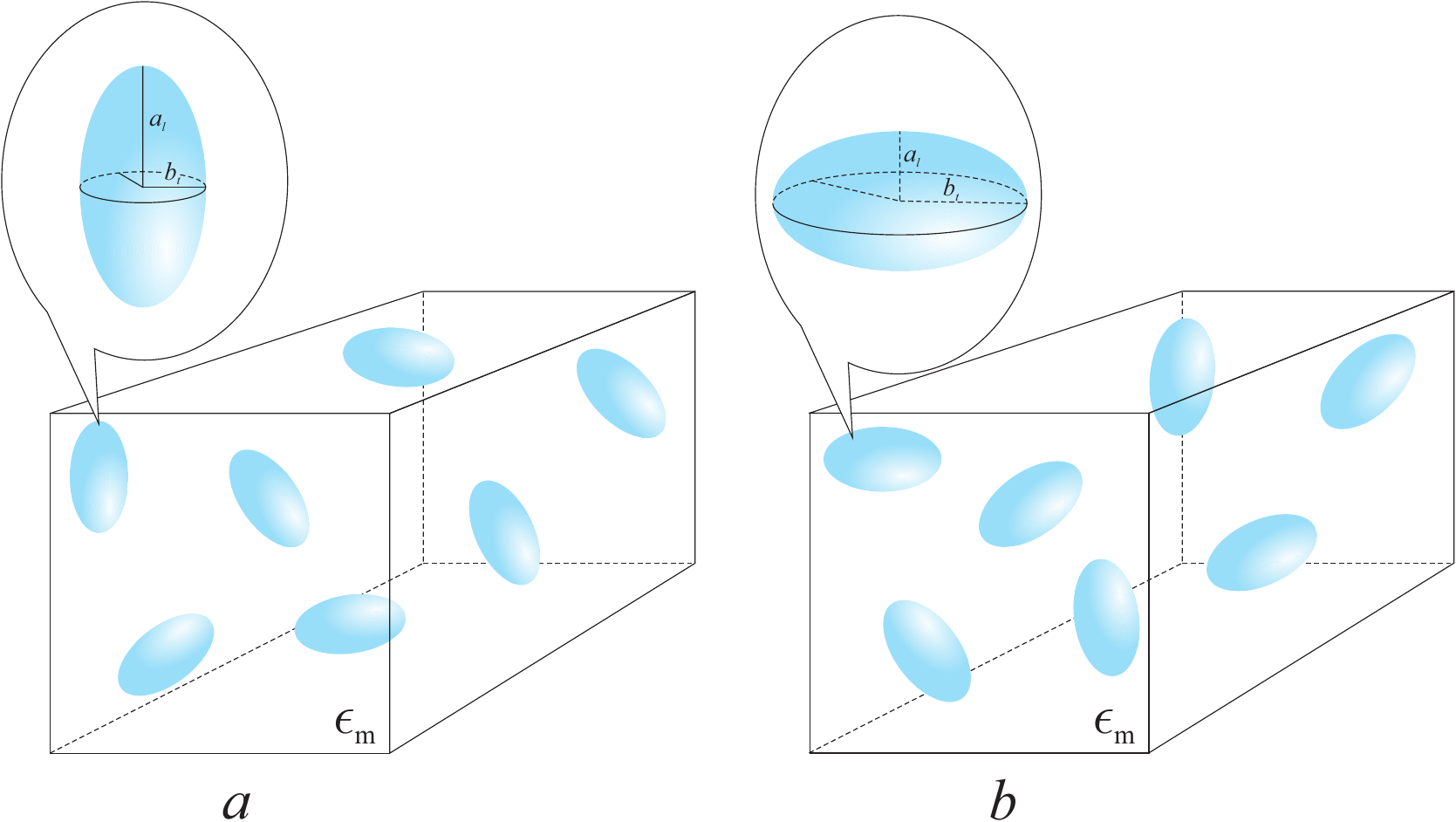}}
\caption{(Colour online) Images of ensembles of prolate (\emph{a}) and oblate (\emph{b}) metallic nanospheroids in a medium with permittivity $\epsilon_{\rm{m}}$.} \label{fig1}
\end{figure}

Let us consider the ensemble of spheroidal metallic nanoparticles with chaotic orientation located in the dielectric medium with the permittivity $\epsilon_{\rm{m}}$ (figure~\ref{fig1}).

We use the statistical averaging \cite{B44} in order to study the optical properties of such an ensemble. In this case, we can investigate the absorption and scattering of a certain ``effective'' particle, whose optical characteristics are statistical in nature. Let us note that this approach has the following advantage: when averaging over the orientations and sizes, the fine-structure details of the absorption and scattering spectra are smoothed, which significantly simplifies the calculations.

The expressions for the ensemble-averaged cross-sections for the absorption and scattering of electromagnetic radiation have the form \cite{B44}

\begin{align}
\label{eq1}
&\left\langle {C_{\rm{abs}}} \right\rangle  = \frac{\omega }{c}\sqrt {{\epsilon_{\rm{m}}}} \left\langle {\Im \alpha } \right\rangle,\\
\label{eq2}
&\left\langle {C_{\rm{sca}}} \right\rangle  = \frac{1}{{6\piup }}\frac{{{\omega ^4}}}{{{c^4}}}\epsilon_{\rm{m}}^2\left\langle {{{\left| \alpha  \right|}^2}} \right\rangle,
\end{align}
where $\omega$ and $c$ are the frequency and rate of the electromagnetic wave, and for the prolate and oblate spheroidal particles
\begin{align}
\label{eq3}
&\left\langle {\Im \alpha } \right\rangle  = \frac{1}{3}\Im \left( {2{\alpha _ \bot } + {\alpha _\parallel }} \right),\\
\label{eq4}
&\left\langle {{{\left| \alpha  \right|}^2}} \right\rangle  = \frac{1}{3}\left( {2{{\left| {{\alpha _ \bot }} \right|}^2} + {{\left| {{\alpha _\parallel }} \right|}^2}} \right),
\end{align}
and the diagonal components of the polarizability tensor
\begin{align}
\label{eq5}
{\alpha _{ \bot \left( \parallel  \right)}}\left( \omega  \right) = V\frac{{{\epsilon_{ \bot \left( \parallel  \right)}}\left( \omega  \right) - {\epsilon_{\rm{m}}}}}{{{\epsilon_{\rm{m}}} + {{\cal L}_{ \bot \left( \parallel  \right)}}\left( {{\epsilon_{ \bot \left( \parallel  \right)}}\left( \omega  \right) - {\epsilon_{\rm{m}}}} \right)}}.
\end{align}
In formula (\ref{eq5}) the volume of spheroid is
\begin{align}
\label{eq6}
V = \frac{4}{3}\piup {a_l}b_t^2,
\end{align}
where $a_l$ and $b_t$ are the major/minor and minor/major semiaxes of prolate/oblate spheroid; ${{\cal L}_{ \bot \left( \parallel  \right)}}$ are the depolarization factors of prolate/oblate spheroids
\begin{align}
\label{eq7}
&\mathcal{L}_{\parallel} = \frac{1 - e_p^2}{2e_p^3}\left( \ln \frac{1 + e_p}{1 - e_p} - 2e_p \right), \quad \mathcal{L}_{\perp} = \frac{1}{2}\left( 1 - \mathcal{L}_{\parallel} \right), \\
\label{eq8}
&{{\cal L}_\parallel } = \frac{{\sqrt {1 - e_p^2} }}{{e_p^3}}\left( {\frac{{{e_p}}}{{\sqrt {1 - e_p^2} }} - {\mathop{\rm arctg}\nolimits} \frac{{{e_p}}}{{\sqrt {1 - e_p^2} }}} \right), \quad {{\cal L}_ \bot } = \frac{1}{2}\left( {1 - {{\cal L}_\parallel }} \right),
\end{align}
where ${e_p}$ are the eccentricities of prolate / oblate spheroid, which are correspondingly equal to

\begin{equation}
\label{eq9}
{e_p} = \sqrt {1 - \frac{{b_t^2}}{{a_l^2}}}, \quad {e_p} = \sqrt {1 - \frac{{a_l^2}}{{b_t^2}}},
\end{equation}
and the diagonal components of the dielectric tensor according to Drude model
\begin{equation}
\label{eq10}
{\epsilon^{ \bot \left( \parallel  \right)}}\left( \omega  \right) = \epsilon_1^{ \bot \left( \parallel  \right)}\left( \omega  \right) + {\mathop{\rm i}\nolimits} \epsilon_2^{ \bot \left( \parallel  \right)}\left( \omega  \right) = {\epsilon^\infty } - \frac{{\omega _p^2}}{{\omega \left( {\omega  + {\mathop{\rm i}\nolimits} \gamma _{{\rm{eff}}}^{ \bot \left( \parallel  \right)}} \right)}}.
\end{equation}
In formula (\ref{eq10}), $\epsilon^\infty$ is the contribution of the interband transitions into the dielectric function; ${\omega _p}$ is the plasma frequency, and the transverse/longitudinal component of the effective relaxation rate  is
\begin{equation}
\label{eq11}
\gamma _{\rm{eff}}^{ \bot \left( \parallel  \right)} = {\gamma _{\rm{bulk}}} + \gamma _{\rm{s}}^{ \bot \left( \parallel  \right)} + \gamma _{\rm{rad}}^{ \bot \left( \parallel  \right)},
\end{equation}
where ${\gamma _{\rm{bulk}}} = {\mathop{\rm const}\nolimits}$ is the bulk relaxation rate, and the transverse (longitudinal) surface relaxation rate and the radiation damping rate are determined by the expressions \cite{B40,B45}
\begin{align}
\label{eq12}
&\gamma_{\rm{s}}^{ \bot \left( \parallel  \right)} = \frac{9}{{16}}\frac{{{{\cal L}_{ \bot \left( \parallel  \right)}}}}{{{\epsilon_{\rm{m}}} + {{\cal L}_{ \bot \left( \parallel  \right)}}\left( {1 - {\epsilon_{\rm{m}}}} \right)}}{\nu _{s,\, \bot }}{\left( {\frac{{{\omega _p}}}{\omega }} \right)^2}\mathscr{F}_{ \bot \left( \parallel  \right)}\left( {{e_p}} \right), \\
\label{eq13}
&\gamma _{\rm{rad}}^{ \bot \left( \parallel  \right)} = \frac{V}{{8\piup }}\frac{{{{\cal L}_{ \bot \left( \parallel  \right)}}}}{{\sqrt {{\epsilon_{\rm{m}}}\left( {{\epsilon^\infty } + \frac{{1 - {{\cal L}_{ \bot \left( \parallel  \right)}}}}{{{{\cal L}_{ \bot \left( \parallel  \right)}}}}{\epsilon_{\rm{m}}}} \right)} }}{\nu _{s,\, \bot }}{\left( {\frac{{{\omega _p}}}{c}} \right)^3}{\left( {\frac{{{\omega _p}}}{\omega }} \right)^2}\mathscr{F}_{ \bot \left( \parallel  \right)}\left( {{e_p}} \right),
\end{align}
where ${\nu _{s,\, \bot }} = {v_{\rm{F}}}/{2{b_t}}$ is the frequency of  individual oscillations of electrons in the transverse direction, ${v_{\rm{F}}}$ is Fermi electron velocity.

The size-dependent functions $\mathscr{F}_{ \bot \left( \parallel  \right)}\left( {{e_p}} \right)$ in (\ref{eq12}), (\ref{eq13}) have the form:

\noindent for prolate spheroids
\begin{align}
\label{eq14}
&\mathscr{F}_ {\bot }\left( {{e_p}} \right) = \frac{1}{{e_p^3}}\left\{ {{e_p}\sqrt {1 - e_p^2} \left( {\frac{1}{2} + e_p^2} \right) + 2\left( {e_p^2 - \frac{1}{4}} \right)\left( {\frac{\piup }{2} - \arcsin \sqrt {1 - e_p^2} } \right)} \right\}, \\
\label{eq15}
&\mathscr{F}_\parallel \left( {{e_p}} \right) = \frac{1}{{e_p^3}}\left\{ {\frac{\piup }{2} - \arcsin \sqrt {1 - e_p^2}  + {e_p}\sqrt {1 - e_p^2} \left( {2 + e_p^2} \right)} \right\},
\end{align}
for oblate spheroids
\begin{align}
\label{eq16}
&\mathscr{F}_ {\bot }\left( {{e_p}} \right) = \frac{{\sqrt {1 - e_p^2} }}{{2e_p^3}}\left[ {\frac{{{e_p}( {1 + e_p^2} )}}{{1 - e_p^2}} + \frac{1}{2}( {1 + 3e_p^2} )\ln \frac{{1 + {e_p}}}{{1 - {e_p}}}} \right], \\
\label{eq17}
&\mathscr{F}_{\parallel }\left( {{e_p}} \right) = \frac{{{{( {1 - e_p^2} )}^{{3 \mathord{\left/
 {\vphantom {3 2}} \right.
 \kern-\nulldelimiterspace} 2}}}}}{{e_p^3}}\left[ {\frac{{{e_p}( {1 + e_p^2} )}}{{{{\left( {1 - e_p^2} \right)}^2}}} - \frac{1}{2}\ln \frac{{1 + {e_p}}}{{1 - {e_p}}}} \right].
\end{align}

By expanding on relations (\ref{eq1}) and (\ref{eq2}) taking into account equations (\ref{eq3})--(\ref{eq5}), we obtain final expressions for the absorption and scattering cross-sections of the ensembles of spheroidal metallic nanoparticles
\begin{align}
\label{eq18}
\left\langle {C_{\rm{abs}}} \right\rangle  ={}& \frac{1}{3}\frac{\omega }{c}\epsilon_{\rm{m}}^{{3 \mathord{\left/
 {\vphantom {3 2}} \right.
 \kern-\nulldelimiterspace} 2}}V \left\{ \frac{{2\epsilon_2^ \bot }}{{{{\left[ {{{\cal L}_ \bot }\epsilon_1^ \bot  + \left( {1 - {{\cal L}_ \bot }} \right){\epsilon_{\rm{m}}}} \right]}^2} + {\cal L}_ \bot ^2{{\left( {\epsilon_2^ \bot } \right)}^2}}}\right.  \nonumber\\
& \left. {}+\frac{{\epsilon_2^\parallel }}{{{{\left[ {{{\cal L}_\parallel }\epsilon_1^\parallel  + \left( {1 - {{\cal L}_\parallel }} \right){\epsilon_{\rm{m}}}} \right]}^2} + {\cal L}_\parallel ^2{{\left( {\epsilon_2^\parallel } \right)}^2}}} \right\},\\
\label{eq19}
\left\langle {C_{\rm{sca}}} \right\rangle  ={}& \frac{1}{{6\piup }}\frac{{{\omega ^4}}}{{{c^4}}}{V^2}\epsilon_{\rm{m}}^2
\left\{ \frac{2}{3}\frac{{{{\left[ {{{\cal L}_ \bot }\left( {{{\left( {\epsilon_1^ \bot  - {\epsilon_{\rm{m}}}} \right)}^2} + {{\left( {\epsilon_2^ \bot } \right)}^2}} \right) + {\epsilon_{\rm{m}}}\left( {\epsilon_1^ \bot  - {\epsilon_{\rm{m}}}} \right)} \right]}^2}}}{{{{\left\{ {{{\left[ {{{\cal L}_ \bot }\epsilon_1^ \bot  + \left( {1 - {{\cal L}_ \bot }} \right){\epsilon_{\rm{m}}}} \right]}^2} + {\cal L}_ \bot ^2{{\left( {\epsilon_2^ \bot } \right)}^2}} \right\}}^2}}}\right.  \nonumber\\
&\left. {}+\frac{1}{3}\frac{{{{\left[ {{{\cal L}_\parallel }\left( {{{\left( {\epsilon_1^\parallel  - {\epsilon_{\rm{m}}}} \right)}^2} + {{\left( {\epsilon_2^\parallel } \right)}^2}} \right) + {\epsilon_{\rm{m}}}\left( {\epsilon_1^\parallel  - {\epsilon_{\rm{m}}}} \right)} \right]}^2}}}{{{{\left\{ {{{\left[ {{{\cal L}_\parallel }\epsilon_1^i + \left( {1 - {{\cal L}_\parallel }} \right){\epsilon_{\rm{m}}}} \right]}^2} + {\cal L}_\parallel ^2{{\left( {\epsilon_2^\parallel } \right)}^2}} \right\}}^2}}} \right\},
\end{align}
where $\epsilon_{1,\,2}^{ \bot \left( \parallel  \right)}$ are the real and imaginary parts of diagonal components of the dielectric tensor.

It should be pointed out that in the limiting cases ${{\cal L}_ \bot } = \frac{1}{2},\,\,{{\cal L}_\parallel } = 0$ (cylinder), ${{\cal L}_ \bot } = 0,\,\,{{\cal L}_\parallel } = 1$ (disk) and ${{\cal L}_ \bot } = {{\cal L}_\parallel } = \frac{1}{3}$ (sphere), the ratios  are obtained for the averaged absorption and scattering cross-sections for the ensembles of the cylindrical, disc-shaped, and spherical nanoparticles, as presented in the paper \cite{B39}.

We investigate the frequency dependencies of the absorption ${\cal K}$, scattering ${\cal S}$, and reflection ${{\cal R}_\infty }$ coefficients using the Kubelka--Munk two-stream approximation \cite{B46}, which is based on the following assumptions:

\begin{itemize}
  \item the medium consists of the flat layer of the finite thickness and the infinite length and width; therefore, no boundary effects occur;
  \item the incident and scattered light fluxes are uniform;
  \item the polarization and spontaneous emission (fluorescence) can be neglected;
  \item the medium is homogeneous and isotropic; any inhomogeneities it contains are small compared to the thickness of the layer;
  \item light is not reflected from the surfaces;
  \item the absorption and scattering coefficients do not depend on the layer thickness.
\end{itemize}

In the described approximation, the relation for the reflection coefficient has the form \cite{B40}:
\begin{equation}
\label{eq20}
{{\cal R}_\infty } = 1 + \frac{{\cal K}}{{\cal S}} - \sqrt {\frac{{{{\cal K}^2}}}{{{{\cal S}^2}}} + 2\frac{{\cal K}}{{\cal S}}},
\end{equation}
where
\begin{equation}
\label{eq21}
{\cal K} = \frac{{\left\langle {{C_{{\rm{abs}}}}} \right\rangle }}{V},
\end{equation}
\begin{equation}
\label{eq22}
{\cal S} = \frac{3}{8}\frac{{\left\langle {{C_{{\rm{sca}}}}} \right\rangle }}{V}\left( {1 - \overline {\cos \theta } } \right),
\end{equation}
and one can assume that $\overline {\cos \theta }  = {1 \mathord{\left/
 {\vphantom {1 2}} \right.
 \kern-\nulldelimiterspace} 2}$.

In order to obtain the numerical results, one uses the formulas (\ref{eq18})--(\ref{eq22}) taking into account the relations (\ref{eq7})--(\ref{eq17}).

\section{Results and discussion}

The calculations were performed for the ensemble of the nanoparticles of different metals in the form of prolate and oblate spheroids with different lengths of the semiaxes (varying eccentricity values). The values of the metal parameters, required for the calculations. are given in table~\ref{tb1}.

\begin{table}[htb]
\caption{Parameters of metals (see, for example, \cite{B45} and references therein).}
\label{tb1}
\begin{center}
\renewcommand{\arraystretch}{1}
\begin{tabular}{|c|c|c|c|c|}
 \hline%
 & \multicolumn{4}{c|}{Parameter}\\%
\cline{2-5}\raisebox{2ex}[0cm][0cm]{Metal}&${r_s}/{a_0}$&${m^*}/{m_e}$&${\epsilon^\infty }$&${\gamma _{{\rm{bulk}}}},\,\,{10^{13}}\,\,{{\rm{s}}^{ - 1}}$\\ \hline%
Pd &4.00&0.37&2.52&13.9\\ \hline%
Pt &3.27&0.54&4.42&10.52\\ \hline%
Ag &3.02&0.96&3.70&2.50\\ \hline%
Au &3.01&0.99&9.84&3.45\\ \hline
Cu &2.11&1.49&12.03&3.70\\ \hline
\end{tabular}
\renewcommand{\arraystretch}{1}
\end{center}
\end{table}

\begin{figure}[htb]
\centerline{\includegraphics[width=0.85\textwidth]{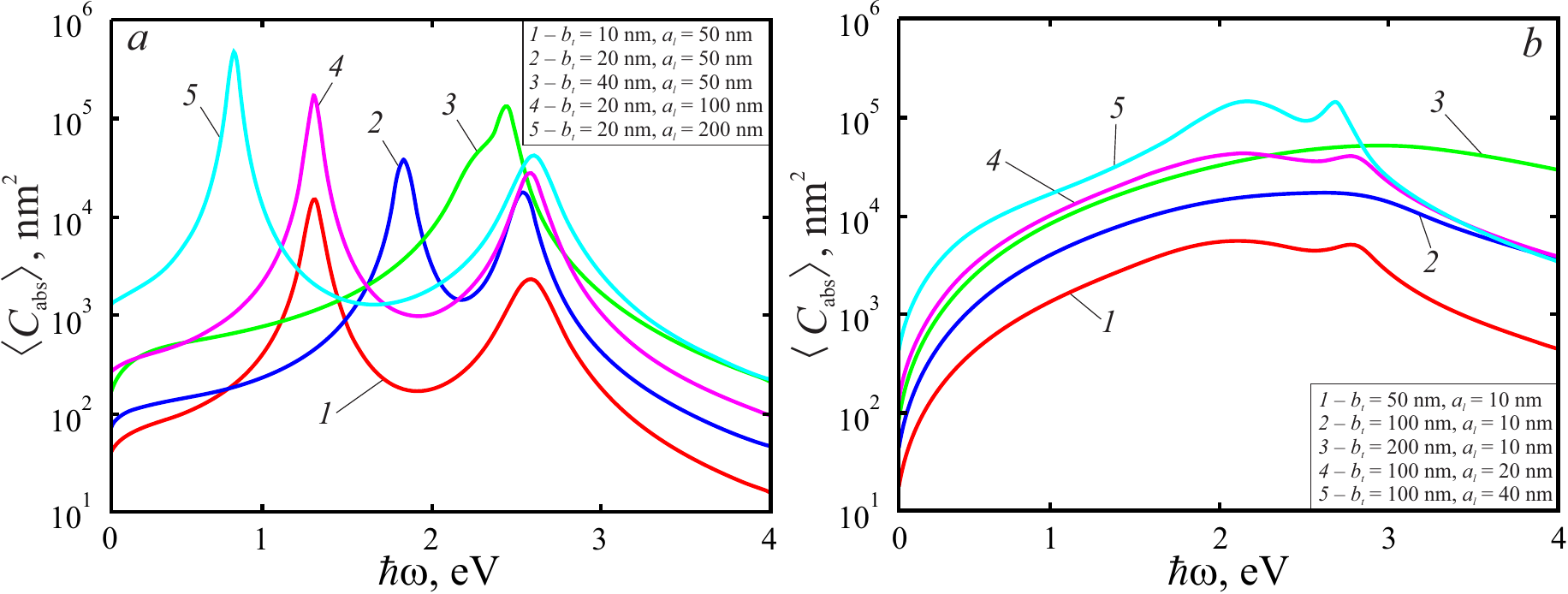}}
\caption{(Colour online) Frequency dependences of averaged absorption cross-sections of assemblies of prolate (\emph{a}) and oblate (\emph{b}) gold spheroids in Teflon with different semi-axes.} \label{fig2}
\end{figure}

The frequency dependencies of the averaged absorption cross-sections for the ensembles of prolate and oblate Au spheroids are shown in figure~\ref{fig2}. The maxima are reached at resonance frequencies corresponding to the zeros of the denominators in relation (\ref{eq5}). For the ensembles of prolate spheroids, the distance between the maxima $\left\langle {{C_{{\rm{abs}}}}} \right\rangle$ is significantly greater than for the ensembles of oblate spheroids, and the maxima are more pronounced. Furthermore, in both cases, as the difference in the semi-axial lengths decreases, the maxima converge (curve 3 in figure~\ref{fig2}\emph{a}). This can be attributed to the fact that the absorption cross-section of the ensemble of the spherical nanoparticles has only one maximum. Let us note that the smaller of the two resonance frequencies for the ensembles of highly prolate spheroids (curves 1, 4 and 5 in figure~\ref{fig2}\emph{a}) lies in the infrared region of the spectrum, whereas in the case of the ensembles of oblate spheroids, all maxima lie in the visible region of the spectrum.

\begin{figure}[htb]
\centerline{\includegraphics[width=0.85\textwidth]{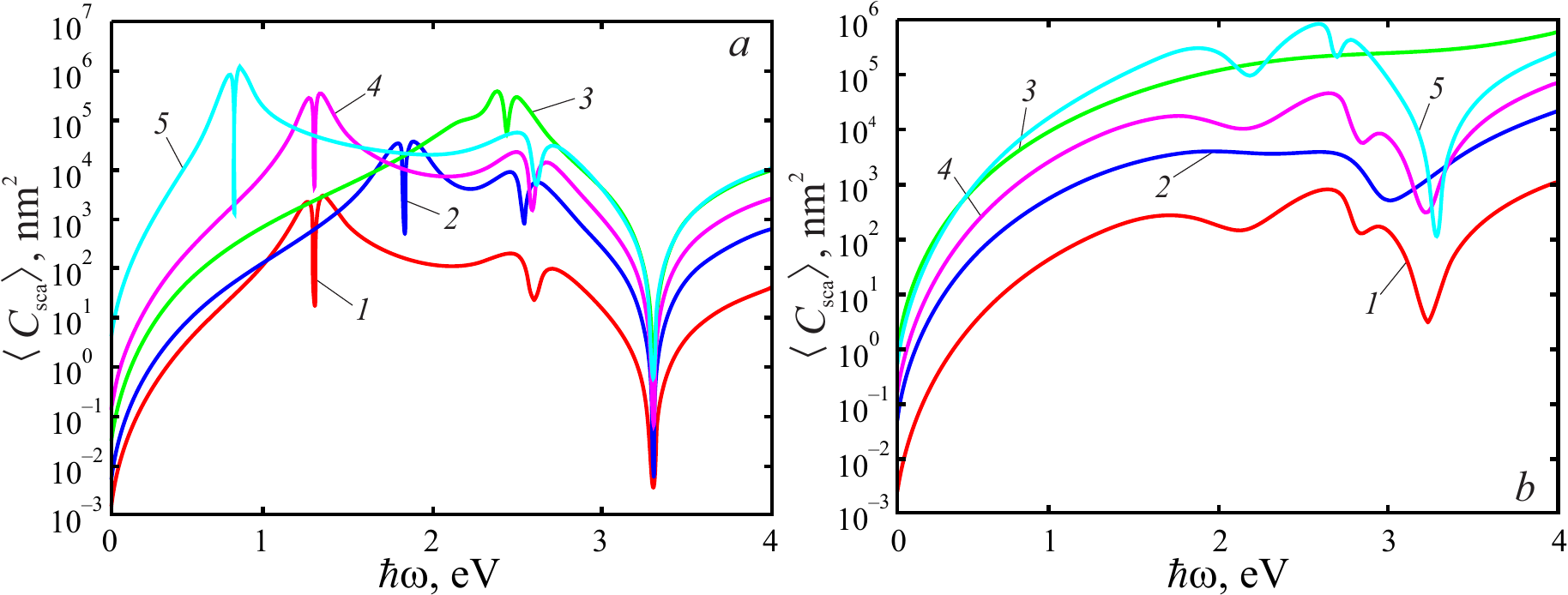}}
\caption{Frequency dependences of the averaged scattering cross-sections of assemblies of prolate (\emph{a}) and oblate (\emph{b}) gold spheroids in Teflon for the same parameter values as in figure~\ref{fig2}.} \label{fig3}
\end{figure}

The frequency dependencies of the averaged scattering cross-sections for the ensembles of prolate and oblate gold spheroids are shown in figure~\ref{fig3}. Let us point out that these dependencies are complex in nature (with a big number of the minima and maxima). This is confirmed by the numerical solution of the equation ${{\rd\langle {{C_{{\rm{sca}}}}}\rangle } / {\rd\omega }} = 0$, which shows the presence of closely spaced roots (table~2 in Appendix~A). Qualitatively, the results for the ensembles of prolate and oblate spheroids are the same (figure~\ref{fig3}). The difference lies in the fact that the amplitude of the maxima for oblate spheroids is from one  to two orders of magnitude smaller. This is due to the difference in the depolarization factors for the prolate and oblate particles. In prolate spheroids, the depolarization factor along the major axis is significantly smaller, which leads to a more substantial local enhancement of the electric field, greater polarizability, and, consequently, more intense plasmon resonance. For oblate spheroids, the depolarization factors are distributed more uniformly, resulting in a weaker field localization and a decrease in the amplitude of the resonance maxima.

The additional reduction in the amplitude is associated with the averaging over the particle orientations. For oblate spheroids, the contribution of the most intense mode after the orientation averaging turns out to be smaller than for prolate spheroids, which leads to a decrease in the maximum values of the averaged scattering cross-sections.

\begin{figure}[htb]
\centerline{\includegraphics[width=0.85\textwidth]{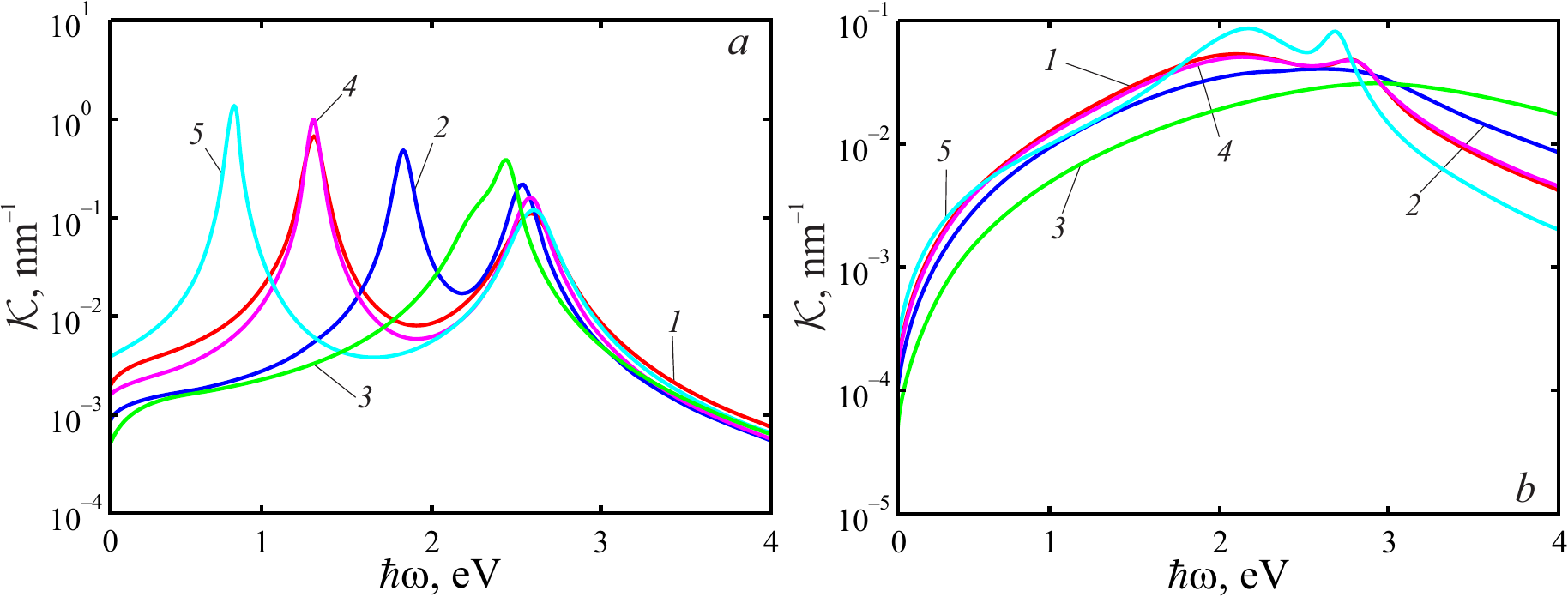}}
\caption{(Colour online) Frequency dependences of the absorption coefficient of assemblies of prolate~(\emph{a}) and oblate (\emph{b}) gold spheroids in Teflon at the same parameter values as in figure~\ref{fig2}.} \label{fig4}
\end{figure}

\begin{figure}[htb]
\centerline{\includegraphics[width=0.85\textwidth]{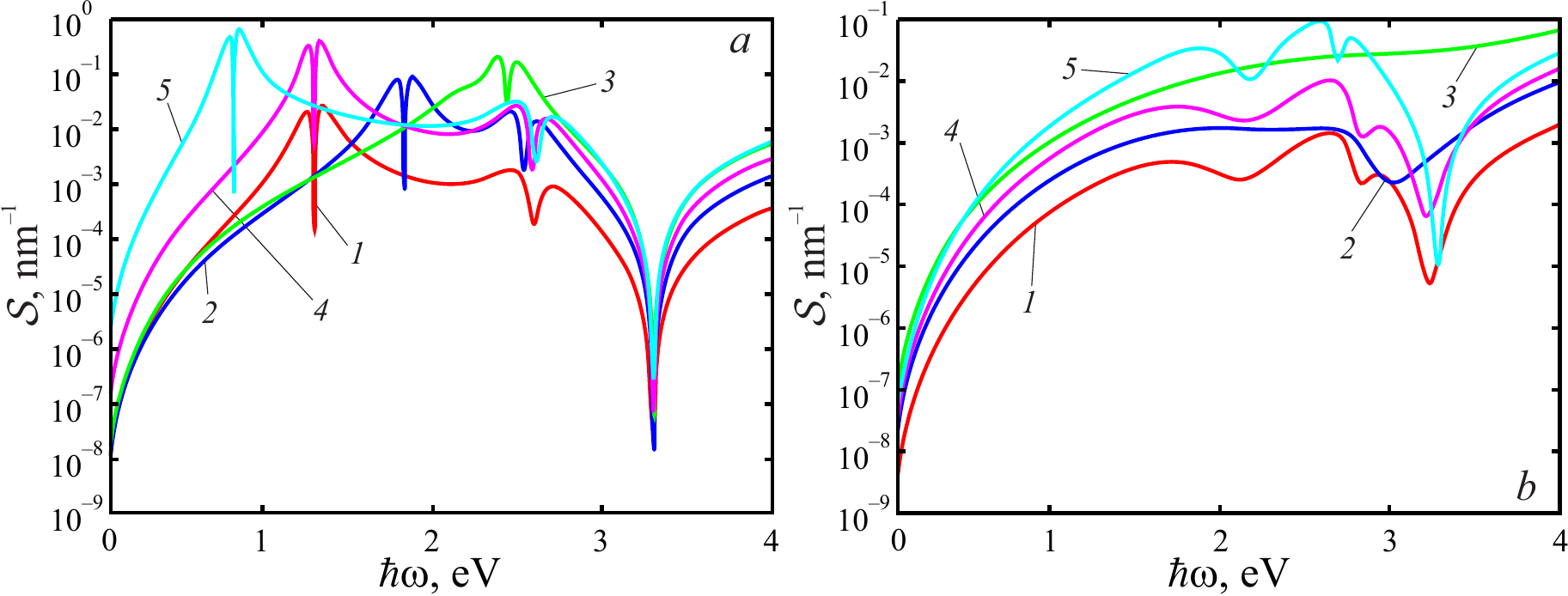}}
\caption{(Colour online) Frequency dependences of the scattering coefficient of assemblies of prolate~(\emph{a}) and oblate (\emph{b}) gold spheroids in Teflon for the same parameter values as in figure~\ref{fig2}.} \label{fig5}
\end{figure}

The frequency dependencies of the absorption and scattering coefficients for the ensembles of gold nanospheroids with different semi-axes are shown in figures~\ref{fig4} and \ref{fig5}. The curves of these dependencies are similar to the curves $\left\langle {C_{\rm{abs}}} \right\rangle \left( {\hbar \omega } \right)$ and $\left\langle {C_{\rm{sca}}} \right\rangle \left( {\hbar \omega } \right)$, and the values ${{\cal K}_{\max }}$ and ${{\cal S}_{\max }}$ for the ensemble of prolate spheroids also exceed the corresponding values for the ensembles of oblate spheroids by one to two orders of magnitude.

\begin{figure}[htb]
\centerline{\includegraphics[width=0.85\textwidth]{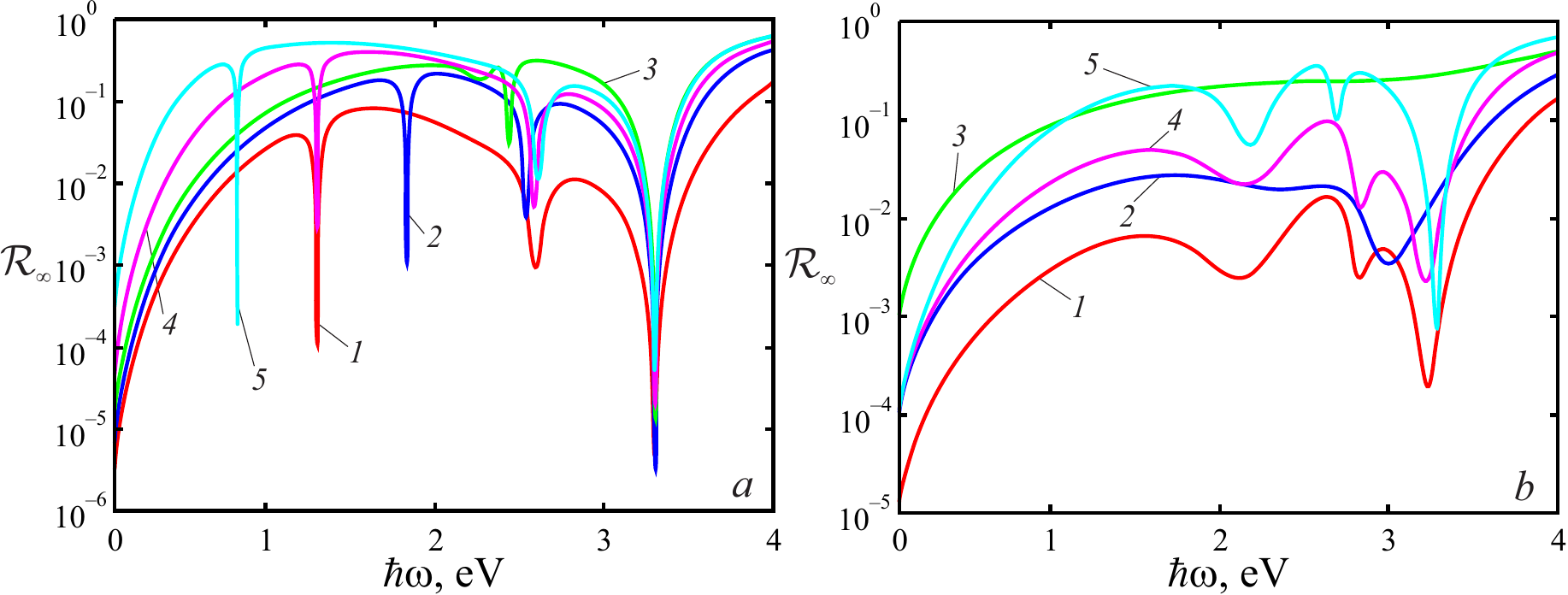}}
\caption{(Colour online) Frequency dependences of the reflection coefficient of assemblies of prolate~(\emph{a}) spheroids   and oblate (\emph{b}) spheroids (${a_l} = 20\,\,{\rm{nm}}$, ${b_t} = 100\,\,{\rm{nm}}$) of different metals in Teflon.} \label{fig6}
\end{figure}

The frequency dependencies of the reflection coefficient for the ensembles of prolate and oblate spheroids made of different metals are shown in figure~\ref{fig6}. The curves for the ensembles of prolate spheroids made of different metals are qualitatively similar; the quantitative differences are determined by the differences in the plasma frequencies and the values of  contributions from  interzone transitions. In the case of the ensembles of oblate spheroids, the curves for  reflection coefficients differ significantly. Thus, for spheroids made of classical plasmonic metals (Ag, Au), the curves of  frequency dependencies exhibit the maxima and minima in the visible frequency range. At the same time, for spheroids made of Cu, Pt, and Pd, the reflection coefficient increases monotoniously in the infrared and visible parts of the spectrum, while the extrema are located in the ultraviolet region of the spectrum. This is due to differences in the complex permittivity of the studied metals, which is determined by the characteristics of their electronic structure. In particular, the positions of the plasmon resonances depend on the plasmon frequency and the spectral position of the interband electronic transitions, which, for Pt and Pd, significantly change the dielectric function even in the visible range, shifting the conditions for the plasmon resonance toward higher energies (shorter wavelengths). For Cu, the interband transitions exert an additional influence, leading to an increase in the optical losses and the shift in the positions of the resonance maxima.

\begin{figure}[htb]
\centerline{\includegraphics[width=0.85\textwidth]{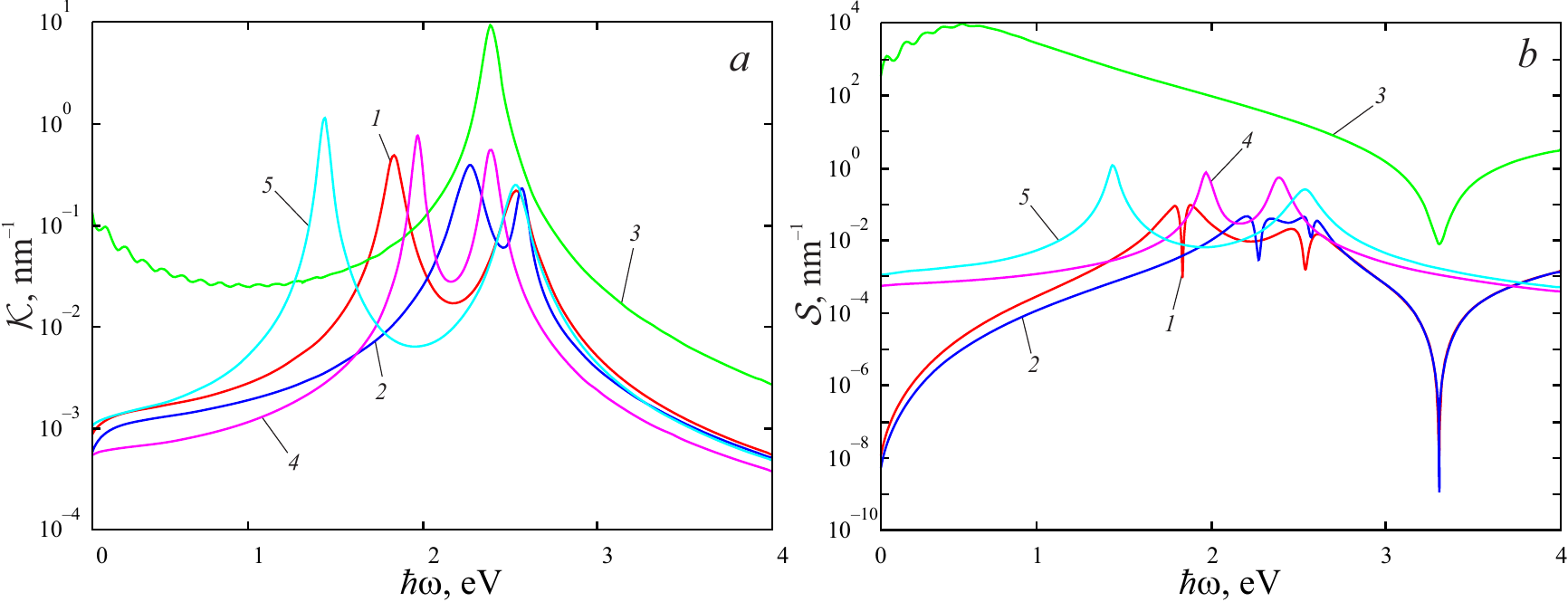}}
\caption{(Colour online) The frequency dependencies for the absorption coefficient~(a) and scattering coefficient (b) of the ensembles of the nanoparticles Au of the different shape and the same volume in Teflon: 1 --- prolate spheroids (${a_l} = 50\,\,{\rm{nm}}$, ${b_t} = 20\,\,{\rm{nm}}$); 2 --- oblate spheroids (${a_l} = 20\,\,{\rm{nm}}$, ${b_t} = 31.6\,\,{\rm{nm}}$); 3 --- spheres ($R = 27.1\,\,{\rm{nm}}$); 4 --- cylinders ($D = 39\,\,{\rm{nm}}$, $H = 70\,\,{\rm{nm}}$); 5 --- disks ($D = 73\,\,{\rm{nm}}$, $H = 20\,\,{\rm{nm}}$).} \label{fig7}
\end{figure}

The comparison of the frequency dependencies of the absorption and scattering coefficients for the ensembles of the nanoparticles of different shapes and identical volumes is shown in figure~\ref{fig7}. It should be pointed out that for the ensembles of the particles of all shapes except spherical ones, the curves ${\cal K}\left( \omega  \right)$ have two maxima, while ${\cal S}\left( \omega  \right)$ have several maxima and minima. Furthermore, the splitting of the absorption coefficient maxima increases in the sequence of the shapes: ``oblate spheroid $\rightarrow$ cylinder $\rightarrow$ prolate spheroid $\rightarrow$ disk''. Since the second $\max \left\{ {{\cal K}\left( \omega  \right)} \right\}$ for the ensembles of the particles of  different shapes are closely spaced and lie in the neighborhood $\max \left\{ {{\cal K}\left( \omega  \right)} \right\}$ of the ensemble of the spherical particles, the first maximum of the absorption coefficient for the ensembles of the disk-shaped particles is reached at the lowest frequency. In turn, the behavior of  frequency dependencies of the scattering coefficient is characterized by the presence of a great number of the maxima and minima, which is analogous to the behavior of the frequency dependencies of the scattering cross-sections.

\section{Conclusions}

The relations for the frequency dependencies of the averaged absorption and scattering cross-sections, as well as the absorption, scattering, and reflection coefficients, for the ensembles of  chaotically oriented metallic nanoparticles in the shape of prolate and oblate spheroids are obtained. The limiting transition to the ensembles of the spherical, cylindrical, and disk-shaped nanoparticles is described.

It is established that the maxima of the averaged absorption cross-section converge as the lengths of the semi-axes of prolate spheroid converge (the particle shape approaches sphere), which is associated with the presence of the single absorption cross-section maximum for the ensemble of  spherical metallic nanoparticles.

It is demonstrated that the shape of  nanoparticles in the ensemble significantly influences the form of the curves of the averaged absorption and scattering cross-sections. For example, the distance between the maxima of the absorption cross-section for the ensemble of prolate spheroids is much greater than for the ensemble of oblate spheroids. The amplitudes of the extrema of the frequency dependencies of the absorption and scattering cross-sections are also greater for the ensemble of prolate spheroids.

It is shown that the complex frequency dependence of the averaged scattering cross-section manifests itself in the presence of a great number of  minima and maxima for the ensembles of spheroids of both types and various sizes, although, for the ensembles of prolate spheroids, the extrema have a greater amplitude.

It is established that the curves of the frequency dependencies of the absorption and scattering coefficients are similar to the corresponding curves of the averaged absorption and scattering cross-sections, which is due to the linear relation between the values, which characterize the identical processes.

It is shown that the transition from prolate spheroids to oblate spheroids in the ensemble results in a decrease in the absorption and in an increase in the scattering.

It is shown that the curves of the frequency dependencies of the reflection coefficient for the ensembles of prolate spheroids made of different metals are qualitatively similar, and that the differences are due to  different contributions from the interband transitions and  different values of the plasma frequency. The corresponding curves for the ensembles of oblate spheroids differ in the location of their extrema: for gold and silver spheroids, the extrema lie in the visible part of the spectrum, whereas for Cu, Pt, and Pd spheroids, they lie predominantly in the ultraviolet region.

The comparison of the frequency dependencies of the absorption and scattering coefficients for the ensembles of the spheroidal nanoparticles and the particles of the limiting shapes (spheres, cylinders and disks) is conducted. The qualitative similarity of the curves for the absorption coefficients is demonstrated (the same number of the maxima for the ensembles of  non-spherical nanoparticles), and quantitative differences are described that are associated with an increase in the splitting of the maxima as the particle shape ranges from oblate spheroid to disk. The study also demonstrates the complex nature of the frequency dependencies of the scattering coefficients, which are qualitatively similar to the corresponding dependencies of the averaged scattering cross-sections of the ensembles of nanoparticles of different shapes.

\section*{Appendix A}

The numerical solutions of the equation ${{\rd\langle {{C_{{\rm{sca}}}}} \rangle }}/{{\rd\omega }} = 0$ as a function of the eccentricity of the spheroidal particles and the corresponding maximum and minimum values of the averaged scattering cross-section are given in table \ref{tb2}.

\begin{table}[h]
\caption{Results of the numerical solution of the equation ${{\rd\langle {{C_{{\rm{sca}}}}} \rangle }}/{{\rd\omega }} = 0$ depending on the eccentricity.}
\label{tb2}
\begin{center}
\renewcommand{\arraystretch}{1.1}
\begin{footnotesize}
\begin{tabular}{|c|c|c|}
 \hline%
 ${e_p}$ & $\hbar {\omega _{{\rm{extr}}}},\,\,{\rm{eV}}$ & $\left\langle {{C_{{\rm{sca}}}}} \right\rangle ,\,\,{\rm{n}}{{\rm{m}}^2}$ \\ \hline
\multirow{8}{*}{$\frac{{\sqrt 5 }}{3}$} & 2.106	& $2.172 \cdot {10^4}$ \\ \cline{2-3}
                                        & 2.176	& $4.913 \cdot {10^3}$ \\ \cline{2-3}
                                        & 2.273	& $2.653 \cdot {10^4}$ \\ \cline{2-3}
                                        & 2.290	& $2.648 \cdot {10^4}$ \\ \cline{2-3}
                                        & 2.409	& $4.028 \cdot {10^4}$ \\ \cline{2-3}
                                        & 2.469	& $3.699 \cdot {10^3}$ \\ \cline{2-3}
                                        & 2.522	& $3.031 \cdot {10^4}$ \\ \cline{2-3}
                                        & 3.302	& $8.752 \cdot {10^{ - 5}}$ \\ \hline
\multirow{8}{*}{$\frac{{\sqrt 3 }}{2}$} & 1.936	& $1.399 \cdot {10^4}$ \\ \cline{2-3}
                                        & 1.991	& 439.452 \\ \cline{2-3}
                                        & 2.047	& $1.468 \cdot {10^4}$ \\ \cline{2-3}
                                        & 2.273	& $3.977 \cdot {10^3}$ \\ \cline{2-3}
                                        & 2.440	& $7.561 \cdot {10^3}$ \\ \cline{2-3}
                                        & 2.511	& 544.928 \\ \cline{2-3}
                                        & 2.574	& $5.426 \cdot {10^3}$ \\ \cline{2-3}
                                        & 3.302	& $1.092 \cdot {10^{ - 3}}$ \\ \hline
\multirow{8}{*}{$\frac{{2\sqrt 2 }}{3}$}& 1.646	& $4.287 \cdot {10^3}$ \\ \cline{2-3}
                                        & 1.694	& 23.028 \\ \cline{2-3}
                                        & 1.745	& $4.884 \cdot {10^3}$ \\ \cline{2-3}
                                        & 2.204	& 371.893 \\ \cline{2-3}
                                        & 2.456	& 726.937 \\ \cline{2-3}
                                        & 2.556	& 58.203 \\ \cline{2-3}
                                        & 2.641	& 447.015 \\ \cline{2-3}
                                        & 3.300	& $1.311 \cdot {10^{ - 3}}$ \\ \hline
\multirow{6}{*}{$\frac{{\sqrt 5 }}{3}$} & 1.821	& $1.246 \cdot {10^5}$ \\ \cline{2-3}
                                        & 2.151	& $4.501 \cdot {10^4}$ \\ \cline{2-3}
                                        & 2.617	& $3.944 \cdot {10^4}$ \\ \cline{2-3}
                                        & 2.733	& $6.722 \cdot {10^4}$ \\ \cline{2-3}
                                        & 2.817	& $1.747 \cdot {10^5}$ \\ \cline{2-3}
                                        & 3.271	& 129.487 \\ \hline
\multirow{6}{*}{${\sqrt 3}$} & 2.006	& $1.112 \cdot {10^6}$ \\ \cline{2-3}
                                        & 2.218	& $2.470 \cdot {10^5}$ \\ \cline{2-3}
                                        & 2.559	& $2.657 \cdot {10^6}$ \\ \cline{2-3}
                                        & 2.637	& $5.158 \cdot {10^5}$ \\ \cline{2-3}
                                        & 2.699	& $1.640 \cdot {10^6}$ \\ \cline{2-3}
                                        & 3.292	& 28.184 \\ \hline
\multirow{8}{*}{$\frac{{\sqrt 5 }}{2}$} & 2.206	& $1.621 \cdot {10^7}$ \\ \cline{2-3}
                                        & 2.284	& $1.355 \cdot {10^6}$ \\ \cline{2-3}
                                        & 2.367	& $1.610 \cdot {10^6}$ \\ \cline{2-3}
                                        & 2.431	& $1.455 \cdot {10^7}$ \\ \cline{2-3}
                                        & 2.505	& $1.936 \cdot {10^7}$ \\ \cline{2-3}
                                        & 2.548	& $3.543 \cdot {10^6}$ \\ \cline{2-3}
                                        & 2.585	& $1.441 \cdot {10^7}$ \\ \cline{2-3}
                                        & 3.302	& 0.464 \\ \hline
\end{tabular}
\end{footnotesize}
\renewcommand{\arraystretch}{1}
\end{center}
\end{table}

\newpage


%
%

\ukrainianpart

\title{Поглинання та розсіювання світла ансамблем сфероїдальних металевих наночастинок}
\author{Н. І.~Павлище\refaddr{label1}, А. В.~Коротун\refaddr{label1,label2}, В. П.~Курбацький\refaddr{label1}}
\addresses{
\addr{label1} Національний університет ``Запорізька політехніка'', Запоріжжя 69011, Україна,
\addr{label2} Інститут металофізики ім. Г.В. Курдюмова НАН України, Київ 03142, Україна,
}
%
%
%

\makeukrtitle

\begin{abstract}
\tolerance=3000%
В роботі досліджено оптичні властивості ансамблів металевих витягнутих та сплюснених наносфероїдів.
Отримано частотні залежності усереднених перерізів поглинання та розсіювання, коефіцієнтів поглинання,
розсіювання та відбивання. Розрахунки проводилися для ансамблів сфероїдальних частинок різних металів.
Відмінність результатів розрахунку частотних характеристик перерізів поглинання та розсіювання для
ансамблів витягнутих та сплюснених сфероїдів виявилася у різному розташуванні та величині максимумів.
Встановлено якісну подібність частотних залежностей коефіцієнта поглинання та розсіювання. Показано,
що частотні криві ансамблів сфероїдів із різних металів суттєво відрізняються розташуванням
екстремумів у межах від видимої області спектра до ультрафіолетової.
\keywords ансамбль наночастинок, витягнуті та сплюснені сфероїди, перерізи поглинання та розсіювання, коефіцієнти поглинання, розсіювання та відбивання, діелектричний тензор

\end{abstract}


\begin{thebibliography}{10}
\bibitem{B1}	Hamanaka Y., Nakamura A., Omi S., Del Fatti N., Vall\'ee F., Flytzanis C., Appl. Phys. Lett., 1999, \textbf{75}, 1712--1714, \doi{10.1063/1.124798}.
\bibitem{B2}	Blaber M. G., Arnold M. D., Harris N., Ford M. J., Cortie M. B., Physica B, 2007, \textbf{394}, 184--187, \doi{10.1016/j.physb.2006.12.011}.
\bibitem{B3}	Pendry J. B., Phys. Rev. Lett., 2000, \textbf{85}, 3966, \doi{10.1103/PhysRevLett.85.3966}.
\bibitem{B4}	Shi L., Gao L., He S., Li B., Phys. Rev. B, 2007, \textbf{76}, 045116, \doi{10.1103/PhysRevB.76.045116}.

\bibitem{B5}    Fang L., Zhong Z., Zhang Y., Huang R., Zhang B., Opt. Express, 2018, \textbf{26}, 30085--30099, \doi{10.1364/OE.26.030085}.
\bibitem{B6}	Ruda H. E., Shik A., J. Appl. Phys., 2007, \textbf{101}, 034312, \doi{10.1063/1.2434971}.

\bibitem{B7}	Rangel-Rojo R., McCarthy J., Bookey H.~T., Kar A.~K., Rodriguez-Fernandez L., Cheang-Wong J.~C., Crespo-Sosa~A., Lopez-Suarez A., Oliver A., Rodriguez-Iglesias V., Silva-Pereyra H.~G., Opt. Commun., 2009, \textbf{282}, 1909--1912, \doi{10.1016/j.optcom.2009.01.048}.
\bibitem{B8}	Lamarre J.-M., Billard F., Kerboua C. H., Lequime M., Roorda S., Martinu L., Opt. Commun., 2008, \textbf{281}, 331--340, \doi{10.1016/j.optcom.2007.09.018}.
\bibitem{B9}	Rangel-Rojo R., Reyes-Esqueda J.~A., Torres-Torres C., Oliver A., Rodriguez-Fernandez L., Crespo-Sosa A., Cheang-Wong J.C., McCarthy J., Bookey H.~T., Kar A.~K., 
In: Silver Nanoparticles, Pozo D. (Ed.), InTech, 2010, \doi{10.5772/8514}.
\bibitem{B10}   Matsuda S.-I., Yasuda Y., Ando S., Adv. Mater., 2005, \textbf{17}, 2221--2224, \doi{10.1002/adma.200500655}.
\bibitem{B12}   Warth A., Lange J., Graener H., Seifert G., J. Phys. Chem. C, 2011, \textbf{115}, 23329--23337, \doi{10.1021/jp2091279}.
\bibitem{B13}	Stalmashonak A., Podlipensky A., Seifert G., Graener H., Appl. Phys. B, 2009, \textbf{94}, 459--465, \doi{10.1007/s00340-008-3309-7}.
\bibitem{B14}	Stookey S. D., Araujo R. J., Appl. Opt., 1968, \textbf{7}, 777--779, \doi{10.1364/AO.7.000777}.
\bibitem{B14+}  Oliver A., Reyes-Esqueda J. A., Cheang-Wong J. C., Rom\'an-Vel\'azquez C. E., Crespo-Sosa A., Rodr\'iguez-Fern\'andez L., Seman J. A., Noguez C., Phys. Rev. B, 2006,
                 \textbf{74}, 245425, \doi{10.1103/PhysRevB.74.245425}.
\bibitem{B15}	Toderas F., Baia M., Baia L., Astilean S., Nanotechnology, 2007, \textbf{18}, 255702, \doi{10.1088/0957-4484/18/25/255702}.
\bibitem{B16}	Talley C. E., Jackson J. B., Oubre C., Grady N. K., Hollars C. W., Lane S. M., Huser T. R., Nordlander P., Halas~N.~J., Nano Lett., 2005, \textbf{5}, 1569--1574, \doi{10.1021/nl050928v}.
\bibitem{B17}	McFarland A. D., Van Duyne R. P., Nano Lett., 2003, \textbf{3}, 1057--1062, \doi{10.1021/nl034372s}.
\bibitem{B18}	Pillai S., Catchpole K. R., Trupke T., Green M. A., J. Appl. Phys., 2007, \textbf{101}, 093105, \doi{10.1063/1.2734885}.
\bibitem{B19}	Nakayama K., Tanabe K., Atwater H. A., Appl. Phys. Lett., 2008, \textbf{93}, 121904, \doi{10.1063/1.2988288}.
\bibitem{B20}	Yeh D.-M., Huang C.-F., Chen C.-Y., Lu Y.-C., Yang C. C., Nanotechnology,  2008, \textbf{19}, 345201, \doi{10.1088/0957-4484/19/34/345201}.
\bibitem{B21}	Fujiki A., Uemura T., Zettsu N., Akai-Kasaya M., Saito A., Kuwahara Y., Appl. Phys. Lett., 2010, \textbf{96}, 043307, \doi{10.1063/1.3271773}.
\bibitem{B21+}  Barad H.~N., Ginsburg A., Cohen H., Rietwyk K.~J., Keller D.~A., Tirosh S., Bouhadana Y., Anderson A.~Y., Zaban~A., Adv. Mater. Interfaces, 2016, \textbf{3}, 1500789, \doi{10.1002/admi.201500789}.
\bibitem{B24}   Dossow L., Kessler R., Sperl M., Born P., Appl. Opt., 2021, \textbf{60}, 10160--10167, \doi{10.1364/AO.441093}.
\bibitem{B25}   Fan X., Zheng W., Singh D. J., Light Sci. Appl., 2014, \textbf{3}, e179, \doi{10.1038/lsa.2014.60}.
\bibitem{B26}   Shao E., Tuersun P., Wumaier D., Li S., Abudula A., Nanomaterials, 2024, \textbf{14}, 1603, \doi{10.3390/nano14191603}.
\bibitem{B27}   Grigorchuk N. I., Tomchuk P. M., Phys. Rev. B, 2011, \textbf{84}, 085448, \doi{10.1103/PhysRevB.84.085448}.
\bibitem{B28}   Grigorchuk N. I., J. Appl. Phys., 2012, \textbf{112}, 064306, \doi{10.1063/1.4751020}.
\bibitem{B29}   Grigorchuk N. I., J. Phys. Stud., 2016, \textbf{20}, 1701, \doi{10.30970/jps.20.1701}.
\bibitem{B30}   Grigorchuk N. I., EPL, 2012, \textbf{97}, 45001, \doi{10.1209/0295-5075/97/45001}.
\bibitem{B31}   Grigorchuk N. I., Eur. Phys. J. B, 2014, \textbf{87}, 252, \doi{10.1140/epjb/e2014-50571-8}.
\bibitem{B32}	Scaffardi L. B., Pellegri N., de Sanctis O., Tocho J. O., Nanotechnology, 2005, \textbf{16}, 158, \doi{10.1088/0957-4484/16/1/030}.
\bibitem{B33}	Beyene H. T., Weber J. W., Verheijen M. A., van de Sanden M. C. M., Creatore M., Nano Res., 2012, \textbf{5}, 513--520, \doi{10.1007/s12274-012-0236-z}.
\bibitem{B34}	Xu G., Tazawa M., Jin P., Nakao S., Appl. Phys. A, 2005, \textbf{80}, 1535--1540, \doi{10.1007/s00339-003-2395-y}.
\bibitem{B35}	Beyene H. T., Tichelaar F. D., Peeters P., Kolev I., van de Sanden M. C. M., Creatore M., Plasma Processes Polym., 2010, \textbf{7}, 657--664, \doi{10.1002/ppap.201000003}.
\bibitem{B36}	Lantiat D., Babonneau D., Camelio S., Pailloux F., Denanot M.-F., J. Appl. Phys., 2007, \textbf{102}, 113518, \doi{10.1063/1.2821914}.
\bibitem{B37}	Chen I.-C., Chen Y.-H., Wang Y.-C., Shih M.-H., Appl. Phys. A, 2013, \textbf{112}, 381--386, \doi{10.1007/s00339-012-7404-6}.
\bibitem{B38}	Baba K., Okuno T., Miyagi M., J. Opt. Soc. Am. B, 1995, \textbf{12}, 2372--2376, \doi{10.1364/JOSAB.12.002372}.
\bibitem{B39}	Korotun A.~V., Pavlyshche N.~I., Phys. Met. Metallogr., 2021, \textbf{122}, 941--949, \doi{10.1134/S0031918X21100057}.
\bibitem{B40}	Pavlyshche N.~I., Korotun A.~V., Kurbatsky V.~P., Low Temp. Phys., 2025, \textbf{51}, 127--132, \doi{10.1063/10.0034657}.
\bibitem{B41}	Qu S., Du C., Song Y., Wang Y., Gao Y., Liu S., Li Y., Zhu D., Chem. Phys. Lett., 2002, \textbf{356}, 403--408, \doi{10.1016/S0009-2614(02)00396-2}.
\bibitem{B42}	Rosi N.~L., Mirkin C.~A., Chem. Rev. 2005, \textbf{105}, 1547--1562, \doi{10.1021/cr030067f}.
\bibitem{B43}   Stuart D.~A., Haes A.~J., Yonzon C.~R., Hicks E.~M., Van Duyne R.~P., IEE Proc., Nanobiotechnol., 2005, \textbf{152}, 13--32, \doi{10.1049/ip-nbt:20045012}.
\bibitem{B44}   Dmitruk N. L., Goncharenko A. V., Venger E. F., Optics of Small Particles and Composite Media, Naukova~Dumka, Kyiv, 2009.
\bibitem{B45}	Korotun A.~V., Pavlyshche N.~I., Funct. Mater., 2022, \textbf{29}, 567--575, \doi{10.15407/FM29.04.567}.
\bibitem{B46}	Kubelka P., Munk F., Z. Techn. Phys., 1931, \textbf{12}, 593--601.


\end{thebibliography}
\end{document}